\documentstyle[preprint,aps,prl,floats,amssymb,cite]{revtex}

\def\m{{\cal M}}
\def\mc{{\cal M}_{\rm cl}}
\def\mq{{\cal M}_{\Lambda}}
\def\CM{{\cal M}}
\def\tr{{\rm Tr\, }}

\def\dd{\hbox{\kern0.3em/\kern-0.7em /\kern0.5em}}
\def\lie#1{{\rm Lie}\left( #1 \right)}
\def\gr{G_r}

\begin{document}

\preprint{\vbox{
\hbox{UCSD/PTH 97--13}
\hbox{hep-th/9706075}
}}
\title{Anomaly Matching Conditions in Supersymmetric
Gauge Theories}
\author{Gustavo Dotti and Aneesh V.~Manohar}
\address{Department of Physics, University of California at San Diego,\\
9500 Gilman Drive, La Jolla, CA 92093-0319}
\date{June 1997}
\maketitle
\begin{abstract}
Sufficient conditions are proven for 't Hooft's consistency conditions to  hold
at points in the moduli space of supersymmetric gauge theories.  Known results
for anomaly matching in supersymmetric QCD are rederived as a sample 
application of the results. The results can be used to show that  the anomaly
matching conditions hold for $s$-confining theories.
\end{abstract}
\pacs{PACS}

One important constraint on the moduli space of vacua of supersymmetric gauge
theories~\cite{susy} is that the massless fermions in the low energy theory
should have the same flavor anomalies as the fundamental fields, i.e.\ the
't~Hooft consistency conditions should be satisfied~\cite{thooft}. The
computation of the flavor anomalies at a point in the moduli space can often be
quite complicated.  In this paper we derive some general conditions which
guarantee that  't~Hooft's consistency conditions are satisfied. The results
are applied to supersymmetric QCD, and agree with known results for this case.
They can also be used to show that 't~Hooft consistency conditions are
satisfied for $s$-confining theories~\cite{sconf}. Other applications will be
discussed in a longer publication~\cite{dm2}

The analysis makes use of the result that the classical moduli space $\mc$ of a
supersymmetric gauge theory is the  algebraic quotient $U \dd G$ of the space
$U$ of critical points $W'=0$ of the superpotential $W$ of the fundamental 
theory by $G$,  the complexification of the gauge group
$\gr$~\cite{dm2,LutTay}.  This follows from the fact that a supersymmetric 
vacuum state is contained in a closed $G$-orbit,  a result proved in
Ref.~\cite{LutTay}.  The relation of supersymmetric vacua to closed orbits, and
a more detailed discussion of the construction of $\mc$ from an algebraic
geometry viewpoint will be given in~\cite{dm2}. There are several subtleties in
the construction not discussed in~\cite{LutTay}  which are relevant for anomaly
matching.

The fundamental theory, such as supersymmetric QCD, will be referred to as the
ultraviolet (UV) theory.  The massless degrees of freedom that characterize the
moduli space $\m$ will be referred to as the infrared (IR) theory. We will use
$\phi \in U$ to represent a constant field configuration in the UV theory.
Supersymmetric vacua are characterized by the values of $\hat\phi \in V$, 
where $V$ is a vector space spanned by gauge invariant polynomials $\hat \phi^i(\phi)$
constructed out of the fundamental fields $\phi^i$. The classical moduli space
$\mc$ is an algebraic set in $V$ (see, e.g. \cite{LutTay}).  There is a natural
map $\pi:U \rightarrow \mc$. The $G$-orbit of a point $\phi \in U$ will be
denoted by $G\phi$. The tangent space at a point $p$ in $X$ will be denoted by
$T_pX$, so that $T_{\phi} U$ is the tangent space at $\phi$ in the UV theory,
and $T_{\hat \phi}\CM$ is the tangent space at $\hat\phi$ in the IR theory. The
differential of $\pi$ at $\phi$, $\pi'_\phi$, gives a map from $T_{\phi} U
\rightarrow T_{\hat \phi}\mc$. In the following discussion, all terms (such as
closed, open, dimension, etc.) are in the algebraic geometry sense. We shall
also assume that the complexified  gauge group is a reductive group. For
example in supersymmetric QCD with $N_F =N_C$ and no tree-level
superpotential,  the UV fields are the quarks $Q^{i\alpha}$ and the antiquarks
$\tilde Q_{j\beta}$ that span a vector space $U$ of dimension $2 N_c^2$. The IR
fields are the mesons $M^i_j$, a baryon $B$ and antibaryon $\tilde B$, that
span the vector space $V$ of dimension $N_c^2+2$. The classical moduli space
$\mc$ is the algebraic set $\det M = B  \tilde B$ contained in $V$. The map
$\pi$ takes $\phi=(Q^{i\alpha},\ \tilde Q_{j\beta})$ to $\hat \phi = (M^i_j,\
B,\ \tilde B)$, where
\begin{equation}
M^i_j = Q^{i \alpha} \tilde Q_{j \alpha},\qquad B = \det Q,
\qquad \tilde B = \det \tilde Q.
\end{equation}

The anomaly matching theorem given below requires that the map $\pi'_\phi:
T_\phi U \rightarrow T_{\hat \phi} \mc$ be surjective, and that $\ker \pi'_\phi
= T_\phi\left( G \phi \right)$. Note that since $\pi$ is gauge invariant, one
always has $T_\phi\left( G \phi \right) \subseteq ker \pi'_\phi$. The following
result proved in Ref.~\cite{dm2} establishes sufficient conditions for these
requirements to hold.

\noindent{\bf Theorem I:}\ \ Assume that $G$ is totally broken at $\phi_0$,
i.e.\ $\lie G \phi_0 \cong \lie G$, and that $G\phi_0$ is closed in $U$. Then
$\ker \pi'_{\phi_0}= T_{\phi_0}\left( G \phi_0 \right)=\lie G$, $\pi'_{\phi_0}$
is onto, and $\pi(\phi_0)$ is a smooth point of $\mc$.

\noindent Note that the condition that $G\phi_0$ be closed is equivalent to the
statement that this orbit contains a point satisfying the $D$-flatness
conditions in Wess-Zumino gauge.

The anomaly matching theorem is

\noindent{{\bf Theorem II:}}\ \ Let $\mc$ be the classical moduli space of a
supersymmetric gauge theory with gauge group $G$ and flavor symmetry $F$. It is
assumed that the gauge theory has no gauge or gravitational  anomalies, and the
flavor symmetries have no gauge anomalies. Let $\hat \phi_0 \in \mc$  be a
point in the classical moduli space. Assume there is a point $\phi_0 \in U$ in
the fiber $\pi^{-1}(\pi(\phi_0))$ of $\hat \phi_0$ such that
\begin{description}
\item[(a)] $G$ is completely broken at $\phi_0$, so that $\lie G \phi_0 
\cong \lie G$.
\item[(b)] $\ker \pi^\prime_{\phi_0} = \lie G \phi_0$ and
$\pi^\prime_{\phi_0}$ is surjective.
\end{description}
If a subgroup $H \subseteq F$ is unbroken at $\hat \phi_0$, then the 't~Hooft
consistency conditions for the $F^3$ flavor anomalies and the $F$ gravitational
anomalies are satisfied.

For the purposes of the proof, it is convenient to write the original flavor
symmetry as $F' \times R$, where $R$ is the $R$-symmetry, and $F'$ now contains
only non-$R$ symmetries. We first prove anomaly matching when $H \subseteq F'$,
and then prove consistency for anomalies that include the $R$ symmetry. (Note
that the unbroken $R$ symmetry might be a linear combination of the original
$R$ symmetry and some generator in $F'$.)

Since $H$ is unbroken at $\hat \phi_0$, $\hat \phi_0$ is $H$-invariant
\begin{equation}
\lie H \hat \phi_0 =0.
\end{equation}
The map $\pi : U \rightarrow \mc$ commutes with the flavor symmetries, so
\begin{equation}
0=\lie H \hat \phi_0 = \lie H \left( \pi \left( \phi_0 \right) \right)
=\pi^\prime_{\phi_0}\left( \lie H \phi_0 \right).
\end{equation}
Thus, by (a) and (b)
\begin{equation}
\lie H \phi_0 \subseteq \ker \pi^\prime_{\phi_0} = \lie G \phi_0.
\end{equation}
This implies that given any ${\frak h} \in \lie H$, there is a unique ${\frak g
\left( h \right )} \in \lie G$ such that 
\begin{equation}\label{3.1}
{\frak h} \phi_0 = - {\frak g \left( h \right )} \phi_0,
\end{equation}
where the minus sign is chosen for convenience.  It is straightforward to check
that the map $\lie H \rightarrow \lie G$ given by $\frak h \rightarrow \frak g
\left( h \right )$ is a Lie-algebra homomorphism,
\begin{equation}
\frak g \left( \left[ h_1, h_2 \right] \right ) = \left[ g \left( h_1 \right ),
g \left( h_2 \right ) \right].
\end{equation}
This allows us to define a new ``star'' representation of $\lie H$ in $U$
\begin{equation}
{\frak h}^* \equiv {\frak h} + {\frak g \left( h \right )}.
\end{equation}
Since $\lie G \phi_0 \subseteq \ker \pi_{\phi_0}^\prime$, the new $\lie H$
representation on $T_{\hat \phi_0} \mc$ defined by $\pi'_{\phi_0}{\frak h}^*$
agrees with  the original one. Thus the
$\frak h^*$-anomalies computed at $\hat \phi_0 \in \mc$ are the same as the
$\frak h$-anomalies at the same point.

$\lie G \phi_0$ is an invariant subspace under ${\frak h}^*$, and the
restriction of ${\frak h}^*$ to $\lie G \phi_0$ is the adjoint action by $\frak
g \left( h \right )$.  This can be seen by direct computation. Take any element
${\frak g} \phi_0 \in \lie G \phi_0$. Then
\begin{equation}
{\frak h}^* {\frak g} \phi_0 = {\frak h g} \phi_0 + {\frak g\left( h \right) g}
\phi_0 = {\frak g h} \phi_0 + {\frak g\left( h \right) g} \phi_0 = {\left[ \frak g
\left( h \right), g \right] } \phi_0 = Ad_{\frak g \left( h \right )}\,
{\frak g} \phi_0,
\end{equation}
since the flavor and gauge symmetries commute, and using Eq.~(\ref{3.1}). The
space $U$ can be broken up into the tangent space to the $G$-orbit  $T_{\phi_0}
G\phi_0 \cong \lie G$ and its invariant complement, $C_{\phi_0}$, since
$G$ is reductive. By (b), the map $\pi^\prime_{\phi_0}$ is a bijective linear 
map from $C_{\phi_0}$ to the tangent space $T_{\hat \phi_0} \mc$ of the moduli
space $\mc$ at  $\hat \phi_0$,  and commutes with $\frak h^*$. Thus the action
of $\frak h^*$ on  $C_{\phi_0}$ is equivalent to the action of $\frak h$ on
$T_{\hat \phi_0} \mc$, by the similarity transformation $S$ given by
$\pi^\prime_{\phi_0}$ restricted to $C_{\phi_0}$. One can write
\begin{equation}\label{3.a}
\frak h^* = {\frak h}_{\rm UV} + {\frak g \left( h \right )} =
\left( \begin{array}{cc}
S\, {\frak h}_{\rm IR}\, S^{-1} & 0 \\
0 & Ad_{\frak g \left( h \right )} \\
\end{array} \right),
\end{equation}
where the second form shows the structure of $\frak h^*$ on $U = C_{\phi_0}
\oplus T_{\phi_0} G\phi_0$. The action of $\frak h$ on $U$ has been labeled by
the subscript UV, and the action on the moduli space has been labeled by IR.

One can now compare anomalies in the UV and IR theories using the two different
forms for $\frak h^*$. Since the adjoint representation is real, the
$\left(\frak h^*\right)^3$ flavor anomaly and $\frak h^*$ gravitation anomaly
are equal to the anomalies in the infrared theory. All that remains is the
proof that the $\left( \frak h^* \right)^3$  and $\frak h^*$ anomalies of $U$
equal the ${\frak h}^3$  and $\frak h$ anomalies of $U$. Let
$\frak h^{\rm A,B,C}$ be any three elements of  $\lie H$. Then
\begin{eqnarray}
\tr \frak h^{* \rm A} \left\{ \frak h^{* \rm B}, \frak h^{* \rm C} \right\}
&=& {\phantom +} \tr \frak h^{\rm A}_{\rm UV} \left\{ \frak h^{\rm B}_{\rm UV}, 
\frak h^{\rm C}_{\rm UV} 
\right\} \nonumber \\
&& + \tr \frak g\left(h^{\rm A}\right) 
\left\{ \frak h^{\rm B}_{\rm UV}, \frak h^{\rm C}_{\rm
UV} 
\right\} + {\rm cyclic} \nonumber \\
&& + \tr
\frak h^{\rm A}_{\rm UV} \left\{ \frak g\left(h^{\rm B}\right), 
\frak g\left( h^{\rm C} \right) \right\} + 
{\rm cyclic} \nonumber \\
&&+\tr \frak g\left( h^{\rm A} \right) \left\{ \frak g\left( h^{\rm B} \right), 
\frak g\left( h^{\rm C} \right) \right\}
\end{eqnarray}
The last three lines vanish because the original theory had no gauge and 
gravitational anomalies, and the flavor symmetries have no gauge anomalies.
Thus the $\frak h^3$ and $\left( \frak h^* \right)^3$  anomalies
are the same. Similarly the $\frak h^*$ and $\frak h$ anomalies agree since
$\frak g$ is traceless because there is no gravitational anomaly.  Thus
't~Hooft's consistency condition for the flavor anomalies is satisfied.

We now prove the matching theorem for anomalies involving the
$R$-charge using an argument similar to the one presented above. The $R$-charge
acting on $U$ is given by the matrix $\frak r$.  The $R$-charge is
defined acting on chiral superfields, and so is the charge of the scalar
component. Anomalies are computed using the fermionic components, so it is
convenient to define a new charge $\tilde {\frak r}$ which we will call
fermionic $R$-charge, defined by
\begin{equation}
\tilde {\frak r} = {\frak r} -1.
\end{equation}
The anomaly can be computed by taking traces over the chiral superfields of
$\tilde {\frak r}$.  The reason for making the distinction between $\frak r$
and $\tilde {\frak r}$  is that the map $\pi$ from $U$ to $\mc$ commutes with
$R=\exp \frak r$, but does not commute with $\tilde R = \exp \tilde{\frak r}$.

Assume that $R$ is unbroken at $\hat \phi_0 = \pi\left( \phi_0 \right)$. Then
by an argument similar to that above, it is possible to define a ``star'' 
$R$-charge, $\frak r^*$,
\begin{equation}\label{3.2}
\frak r^* \equiv r + g\left( r \right)
\end{equation}
which has the form
\begin{equation}\label{3.b}
 \frak r^* = r_{\rm UV} + g \left( r \right ) =
\left( \begin{array}{cc}
S\, {\frak r_{\rm IR}}\, S^{-1} & 0 \\
0 & Ad_{\frak g \left( r \right )} \\
\end{array} \right)
\end{equation}
under the decomposition of $U$ into $C_{\phi_0} \oplus T_{\phi_0} G \phi_0$. As
in Eq.~(\ref{3.a}), we have used the subscripts UV and IR to denote the $R$
charges in the ultraviolet and infrared theories. Note that $S$ is the same
matrix in Eqs.~(\ref{3.a},\ref{3.b}), given by $\pi^\prime_{\phi_0}$ restricted
to $C_{\phi_0}$. The fermionic $R$-charge is then given by
\begin{equation}\label{14}
 \tilde {\frak r}^* = {\frak r}^* -1 = \tilde {\frak r}_{\rm UV} + 
 {\frak g \left( r \right )} =
\left( \begin{array}{cc}
S\, \tilde {\frak r}_{\rm IR}\, S^{-1} & 0 \\
0 & Ad_{\frak g \left( r \right )} -1 \\
\end{array} \right)
\end{equation}
where in the last equality we have used the fact that fermion $R$ charge
$\tilde {\frak r}_{\rm IR} = {\frak r}_{\rm IR}-1$ in the infrared theory.

Compute the trace of $(\tilde {\frak r}^*)^3$ in $U$,
\begin{equation}
\tr \left( \tilde {\frak r}^* \right)^3 = \tr \left\{ \tilde {\frak r}_{\rm UV} 
+ {\frak g\left( r \right)} \right\}^3
= \tr \left\{ \tilde {\frak r}^3_{\rm UV} + 3 \tilde {\frak r}^2_{\rm UV} 
{\frak g\left( r \right)} + 3 \tilde {\frak r}_{\rm UV} 
{\frak g\left( r \right)}^2 + {\frak g\left( r \right)}^3 \right\}.
\end{equation}
The $R$-charge has no gauge anomaly, so $\tr_U\,  \tilde {\frak r}_{\rm UV} 
\{{\frak g_{\rm A}, g_{\rm B}}\} + \tr_{\lie G}\, \{Ad_{\frak g_{\rm A}},
Ad_{\frak g_{\rm B}}\} =0$,  for any ${\frak g}_{\rm A,B} \in \lie G$.  Here
the first term is the matter contribution to the anomaly, and the  second term
is the gaugino contribution. The absence of gauge anomalies implies that odd
powers of $\frak g\left( r \right)$ vanish when traced over the matter fields,
since there is no gaugino contribution to these anomalies. Thus we find
\begin{equation}\label{3.5}
\tr_{U} \left( \tilde {\frak r}^* \right)^3 = \tr_{U} 
\left( \tilde {\frak r}_{\rm UV} \right)^3 - 3\, \tr_{\lie G}\,
 Ad_{\frak g\left( r \right)}^2.
\end{equation}
The block diagonal form of $\tilde{\frak r}^*$ Eq.~(\ref{14}), gives
\begin{equation}\label{3.11}
\tr_U \left( \tilde {\frak r}^* \right)^3  = \tr \left( \tilde 
{\frak r}_{\rm IR} \right)^3
-  \tr_{\lie G} \left( 1 + 3 Ad_{\frak g\left( r \right)}^2 \right).
\end{equation}
The $R^3$ anomaly $A_{\rm UV}\left( R^3 \right)$ in the UV theory is given by 
adding the matter and gaugino contributions
\begin{equation}\label{3.10}
A_{\rm UV}\left( R^3 \right)=
\tr_U  \left( \tilde {\frak r}_{\rm UV} \right)^3 + \tr_{\lie G} 1^3 = 
\tr_U \left( \tilde {\frak r}^* \right)^3 + \tr_{\lie G} \left( 1 + 3 Ad_{
\frak g \left( r \right)}^2 \right).
\end{equation}
The $R^3$ anomaly $A_{\rm IR}\left( R^3 \right)$ in the IR theory is given by
\begin{equation}\label{3.12}
A_{\rm IR}\left( R^3 \right) = \tr \left( \tilde {\frak r}_{\rm IR} \right)^3, 
\end{equation}
since there are no gauginos in the low energy theory.  Combining
Eq.~(\ref{3.5}--\ref{3.12}),  one sees immediately that the UV and IR anomalies
are equal, $A_{\rm UV}\left( R^3 \right)=A_{\rm IR} \left( R^3 \right)$.

It is straightforward to check that the gravitational $R$ anomaly, and the $H^2
R$ and $H R^2$ anomalies match by a similar computation;  the details are given
in Ref.~\cite{dm2}.

The results derived above allow one to study the matching of
anomalies between the ultraviolet and infrared theories at certain points in
the classical moduli space. We now derive some results that allow one to relate
the anomalies at different points on the moduli space to each other. The moduli
space is no longer restricted to be the classical moduli space $\mc$. The first
case we will consider is when the moduli space $\CM$ is an algebraic set in
an ambient vector space $V$ given as the critical points of a flavor symmetric
superpotential $W$ with $R$-charge two,
\begin{equation}\label{ir:2}
\CM = \left\{ \hat \phi \in V \, | \, W_i(\hat \phi )=0 \right\},
\end{equation}
where $\hat \phi$ denotes a point in $V$, and we will use the notation
$W_i\equiv \partial W / \partial \hat \phi^i$, $W_{ij}\equiv \partial^2 W /
\partial \hat \phi^i\partial \hat \phi^j$, etc. The tangent space to $\CM$ at
$\hat \phi_0$, $T_{\hat \phi_0} \CM$, is defined by
\begin{equation}\label{ir:1}
T_{\hat \phi_0}\CM = 
\left\{ \hat v^i \in V \, | \, W_{ij} (\hat \phi_0 ) \hat v^j=0 \right\}.
\end{equation}
In all the cases we are interested in, $W$ is a polynomial in $\hat \phi$ and
Eq.~(\ref{ir:1}) agrees with the algebraic geometry notion of the tangent
space.

Assume that a subgroup $H$ (not containing an $R$ symmetry) of the flavor
symmetry group $F$ is unbroken at a point $\hat \phi_0 \in \CM$. The invariance
of the superpotential $W$ under $F$ implies that
\begin{equation}
W \left( h^i_j\, \hat \phi^j \right) = W(\hat \phi^i),
\end{equation}
where $h^i_j$ is the matrix for the $H$ transformation in the representation
$\rho$ of the fields $\hat \phi$. Differentiating this equation twice with respect
to $\hat \phi$ and using $H\hat \phi_0=\hat \phi_0$ gives
\begin{equation}
h^k_i\, h^l_j\, W_{kl} \left(\hat \phi_0 \right) = 
W_{ij} ( \hat \phi_0),
\end{equation}
which shows that $W_{ij}( \hat \phi_0 )$ is a $H$ invariant tensor that
transforms as $\left( \bar \rho \otimes \bar \rho \right)_S$ under $H$.  The
tangent space to $\CM$ at $\hat \phi_0$ is the null-space of $W_{ij}$, and so
is $H$-invariant. One can write $V = T_{\hat \phi_0} \CM + N_{\hat \phi_0}\CM$
as the direct sum of the tangent space and its orthogonal complement in $V$.
Then $W_{ij}$ provides a non-singular invertible map from $N_{\hat \phi_0} \CM$
into its dual, so that $N_{\hat \phi_0} \CM$ transforms as a real
representation of $H$. This immediately implies that the $H$ anomalies computed
using the flat directions  $T_{\hat \phi_0} \CM$ agree with those computed
using the entire vector space $V$.

A similar result holds for the anomalies involving the $R$ charge.  Let $R_i$
be the $R$-charge of $\hat \phi_i$, so that
\begin{equation}
W\left( e^{i \alpha R_i}\, \hat \phi^i \right) = e^{2 i \alpha} \,
W(\hat \phi^i),
\end{equation}
since $W$ has $R$ charge two. Differentiating twice with respect to $\hat \phi$
shows that
\begin{equation}
e^{i \alpha \left( R_i +R_j\right)}\,
W_{ij} (\hat  \phi_0 ) = e^{2 i \alpha }\,
W_{ij} (\hat  \phi_0 ),
\end{equation}
which can be written in the suggestive form
\begin{equation}\label{ir:3}
e^{i \alpha \left( \left[R_i-1\right] + \left[ R_j - 1 \right] \right)}\, 
W_{ij} (\hat  \phi_0 ) =
W_{ij} (\hat  \phi_0).
\end{equation}
$R_i-1$ is the $R$ charge of the fermionic component of the chiral superfield.
Thus Eq.~(\ref{ir:3}) shows that $N_{\hat \phi_0} \CM$ transforms like a real
representation under $\tilde R = R-1$, the fermionic $R$ charge. Thus the $R$
anomalies, (and mixed anomalies involving $R$ and non-$R$ flavor symmetries) 
can be computed at $\hat \phi_0$ using $V$ instead of $T_{\hat \phi_0} \CM$.
The result can be summarized by 

\noindent{\bf Theorem III:} Let $\CM \subseteq V$ be a moduli space described
by the critical points of a flavor symmetric  superpotential $W$ with
$R$-charge two. Then the anomalies of an unbroken subgroup $H \subseteq F$ at a
point $\hat \phi_0 \in \CM$ can be computed using the entire space $V$, instead
of $T_{\hat\phi_0}\CM$. If the anomaly matching conditions between the
UV and IR theories for $H$ are satisfied at $\hat \phi_0$, they are also
satisfied at all points of any moduli space $\CM' \in V$ given by the critical
points of any $W'$ (including $W'=0$ and $W'=W$).

Note that this result tells us that for moduli spaces described by invariant
superpotentials, the precise form of the moduli space is irrelevant. The only
role of possible quantum deformations is to remove points of higher symmetry
from the moduli space. It also greatly simplifies the computation of anomalies
in the infrared theory, since one does not need to compute the tangent vectors
at a given point in the moduli space.

One simple application of the above result is to prove that anomaly matching
conditions are compatible with integrating out heavy fields. Assume that one
has a theory with a moduli space $\mq$ described by a superpotential $W( \hat
\phi, \Lambda)$. Now perturb the UV theory by adding a tree level mass term
$m_{ij} \phi^i \phi^j$ to the superpotential. $m_{ij} \phi^i \phi^j$ is gauge
invariant, and can be written as a polynomial $W_m( \hat \phi)$  of the gauge
invariant composites $\hat \phi$ of the IR theory. If the UV theory contains no
singlets, then $W_m(\hat \phi)$ is linear in the basic gauge invariant
composite fields $\hat \phi$. From this, it immediately follows that the
effective superpotential of the massive theory is given by
\begin{equation}
W(\hat \phi,\Lambda) = W_0(\hat \phi,\Lambda) + W_m(\hat \phi),
\end{equation}
where $W_0$ is the superpotential in the absence of a mass term, since a linear
term in the fields is equivalent to a redefinition of the source.

The anomalies in the IR theory for any unbroken subgroup are unaffected by the
change in the moduli space due to the addition of the mass term. They are still
obtained by tracing over the whole space $V$. In the UV theory, one should
trace not over the whole space $U$, but only over the modes that remain
massless when $W_m$ is turned on. But it is easy to see that the massive modes
in the UV theory form a real representation of the unbroken symmetry. The
argument is the same as that used in the IR theory, except that $W_{ij}(\hat
\phi_0)$ is replaced by the (constant) matrix $m_{ij}$. The mass term does
not introduce any modifications to the anomaly in the UV or IR theory for any
symmetry left unbroken by the mass. Thus one finds that if the 't~Hooft
conditions are verified for a theory with  a moduli space given by a
superpotential, they are also valid for any theory obtained by integrating out
fields by adding a mass term.

One can now apply the results to study anomaly matching in supersymmetric gauge
theories. Consider supersymmetric QCD with $N_F \ge N_c >2$.
The fundamental fields are the quarks $Q^{i\alpha}$ and
antiquarks $\tilde Q_{j\beta}$. The flavor symmetry group is $SU(N_F)_L \times
SU(N_F)_R \times U(1)_B \times U(1)_R$ if $N_c > 2$.\footnote{For $N_c=2$, the
$SU(N_F)_L \times SU(N_F)_R \times U(1)_B$ is enlarged into a $SU(2N_F)$ flavor
symmetry. Anomaly matching can be proven by an argument similar to that for
$N_c >2$.}

Consider the point $\phi_0$ in the UV theory
\begin{equation}
Q^{i\alpha} = \left\{ \begin{array}{ll}
m \delta^{i\alpha} & i \le N_c \\
0 & i > N_c 
\end{array}\right. , \qquad
\tilde Q_{j \alpha} = 0.
\end{equation}
The point $\pi(\phi_0) = \hat \phi_0$ in the IR theory is described by gauge
invariant meson and baryons fields,
\begin{equation}\label{phiM}
M^i_j =0, \qquad \tilde B^{j_1\cdots j_s} =0, 
\qquad B_{i_1 \cdots i_s} =
m^{N_c} \ \epsilon_{12...N_c i_1 \cdots i_s}.
\end{equation}
The unbroken flavor group at $\hat \phi_0$ is $SU(N_c)_L \times
SU(N_F-N_c)_L \times SU(N_F)_R \times U(1)_B \times U(1)_R$. Under these
unbroken symmetries, the fields transform as
\begin{center}
\begin{tabular}{|c|c|c|c|c|c|}
\hline
 & $SU(N_c)_L$ & $SU(N_F-N_c)_L$ & $SU(N_F)_R$ & $U(1)_B$ & $U(1)_R$ \\
\hline
$Q^{i\alpha}$, $i \le N_c$ & $N_c$ & $-$ & $-$ & $0$ & $0$ \\
$Q^{i \alpha}$, $i > N_c$ & $-$ & $N_F - N_c$ & $-$ & $- N_F$ & 
$(3N_F-4N_c)/(2N_F-N_c)$ \\
$\tilde Q_{j \alpha}$ & $-$ & $-$ & $\overline{N}_F$ & $N_F - N_c$ & 
$(3N_F-4N_c)/(2N_F-N_c)$ \\
\hline
\end{tabular}
\end{center}
The unbroken $U(1)_B$ and $U(1)_R$ symmetries are linear combinations of 
the original  $U(1)_B$ and $U(1)_R$ and a $U(1)$ generator in $SU(N_F)_L$.

The point $\phi_0=(Q^{i\alpha}, Q_{j \alpha})$ breaks the gauge group
completely. The orbit $G\phi_0$ is closed and has maximal dimension, so
Theorem~I tells us that the hypotheses of Theorem~II are satisfied.
The anomaly matching theorem (Theorem~II) implies that the $SU(N_c)_L \times
SU(N_F-N_c)_L \times SU(N_F)_R \times U(1)_B \times U(1)_R$ anomalies must
match between the UV and IR theories. It is straightforward to verify by
explicit computation that this is the case. The UV anomalies are computed using
the above transformation rules for the fundamental fields. The IR anomalies are
computed by determining the  representation of the tangent vectors to the
classical moduli space at $\hat \phi_0$ under the unbroken symmetry. One can
similarly show that anomaly matching holds at other points at the moduli space
of supersymmetric QCD. 

In the special case $N_F=N_c+1$, the classical moduli space is described by a
superpotential. Then Theorem~III implies that since the $SU(N_c)_L \times
SU(N_F-N_c)_L \times SU(N_F)_R \times U(1)_B \times U(1)_R$ match at
Eq.~(\ref{phiM}), they must also match at the origin. One can verify by
explicit computation that these anomalies  match at the origin for $N_F=N_c+1$,
but not for any other value of  $N_F$. (Equivalently, the fact that the these
anomalies match at  Eq.~(\ref{phiM}) but not at the origin for $N_F > N_c+1$
implies that  the moduli space for $N_F > N_c+1$ cannot be given by a
superpotential.) Similarly by considering the point given by exchanging the
values of $Q$ and $\tilde Q$ in Eq.~(\ref{phiM}), one can prove that the 
$SU(N_F)_L \times SU(N_c)_R \times SU(N_F-N_c)_R \times U(1)_B \times  U(1)_R$
anomalies also match at the origin. Anomaly matching for these two subgroups is
sufficient to guarantee that the anomalies for the full $SU(N_F)_L \times 
SU(N_F)_R \times U(1)_B \times  U(1)_R$ flavor group match at the origin.
Applying Theorem~III again then shows that the anomalies match everywhere on
the moduli space.

Since supersymmetric QCD with $N_F=N_c+1$ is described by a  superpotential,
integrating out one flavor by adding a mass term also  gives a consistent
theory. This is supersymmetric QCD with $N_F=N_c$ with  the quantum deformed
moduli space $\det M - B \tilde B = \Lambda^{2N_c}.$\footnote{Note that the anomaly
matching theorem cannot be applied at the origin,  which is a  point of the
classical moduli space, but is not part of the quantum  moduli space.}
Integrating out additional flavors leads to a trivial result, since there is no
point in the moduli space of the theory, and all vacua are unstable because of
the quantum superpotential~\cite{ads}. One cannot
relate anomalies in $N_F > N_c+1$ to those for $N_F=N_c+1$ by adding mass
terms, since the theories with $N_F > N_c+1$ are not described by a
superpotential. This is consistent with the result that these theories do not
satisfy the anomaly matching conditions at the origin, and the infrared
behavior is governed by a dual theory~\cite{susy}. 

The results of this paper have been used to reproduce known results for
supersymmetric QCD, without having to explicitly compute any anomalies in the
UV or IR theories. The key point is to find some simple field configurations
$\phi$  that completely break the gauge symmetry, and at  which $\pi'$ is
surjective. The results are particularly powerful for theories with a moduli
space described by a superpotential. The results of this paper can also be
applied to the $s$-confining theories that have  been studied
recently~\cite{sconf}. These theories have a moduli space given by a
superpotential, and are therefore similar to supersymmetric QCD for
$N_F=N_c+1$. The 't~Hooft consistency conditions are automatically satisfied
for the entire moduli space, using Theorems~I--III. It then follows that any
theory obtained from an $s$-confining theory by adding mass terms also
satisfies the 't~Hooft consistency conditions, as long as it has supersymmetric
vacua. A direct check of the anomaly matching conditions by explicit
computation is extremely involved.

We are indebted to N.~Wallach and M.~Hunziker for extensive discussions on 
algebraic geometry, N.~Wallach for providing a draft of
his  book~\cite{GooWall} prior to publication, and
K.~Intriligator for discussions on supersymmetric gauge theories and duality.
This work was supported in part by a Department of Energy grant 
DOE-FG03-90ER40546.

\end{document}